\documentclass[showpacs,showkeys,twocolumn]{revtex4}

\usepackage{amsfonts,amsmath,amssymb,bm,hyperref}

\begin{document}


\title{Neutrino spin relaxation in medium with stochastic characteristics}

\author{Maxim Dvornikov}
\affiliation{Institute of Terrestrial Magnetism, Ionosphere and
\\ Radiowave Propagation (IZMIRAN) \\ 142190, Troitsk, Moscow
region, Russia} \email{maxdvo@izmiran.ru}

\date{\today}

\begin{abstract}
The helicity evolution of a neutrino propagating in randomly
moving and polarized matter is studied. The type of the neutrino
interaction with background fermions is arbitrary. We derive the
equation for the description of the averaged neutrino helicity
evolution. In the particular case of a $\tau$-neutrino interacting
with ultrarelativistic electron-positron plasma we obtain the
expression for the neutrino helicity relaxation rate in the
explicit form. We study the neutrino spin relaxation in the
relativistic primordial plasma. Supposing that the conversion of
left-handed neutrinos into right-handed ones is suppressed at the
early stages of the Universe evolution we get the upper limit on
the $\tau$-neutrino mass.
\end{abstract}

\pacs{14.60.Pq, 52.25.Gj}

\keywords{neutrino helicity, electromagnetic fluctuations,
relativistic plasma, early Universe}

\maketitle

It is well known that if a particle, possessing the magnetic
moment, interacts with the external electromagnetic field,
particle's spin will rotate around a certain direction determined
by this external field. The covariant description of the
particle's spin precession in the external electromagnetic field
was presented in Ref.~\cite{BarMicTel59}. Recently it was
established (see Refs.~\cite{EgoLobStu99,EgoLobStu00}) that weak
interactions can also cause the particle's spin precession. A
neutral $1/2$ spin particle (e.g., a neutrino) interacting with
matter via weak currents was studied in
Refs.~\cite{EgoLobStu99,EgoLobStu00,LobStu01,DvoStu02JHEP}. It was
found that the particle's spin precession rate depends on the
velocities and polarizations of the background matter fermions. On
the basis of the elaborated approach one can study various
interesting phenomena such as, for instance, neutrino spin light
and self-polarization effect in matter (see
Refs.~\cite{LobStu03,LobStu04}).

In case of a neutrino interacting with medium the effect of
velocities and polarizations of the background fermions on the
neutrino's spin rotation is suppressed by the factor $1/\gamma$,
where $\gamma=E_\nu/m_\nu$. Thus for relativistic neutrinos the
transitions between left and right polarized particles due to the
matter effects vanish. In order to achieve the essential neutrino
spin rotation rate one should have background fermions with very
high velocities or polarizations. Such ultrarelativistic matter
can exist mainly in various astrophysical environments or in the
relativistic plasma of the early Universe.

Neutrinos in the early Universe were discussed for the first time
in Ref.~\cite{AlpFolHer53}. Since then a great deal of papers
devoted to this topic have been published. The comprehensive
review of neutrinos in cosmology was presented in
Ref.~\cite{Dol02}. The neutrino mass constraints can also be
obtained from the studying of the early Universe evolution. For
example, one extracts information about neutrino mass from the
Cosmic Microwave Background radiation, power spectrum in large
scale structure surveys as well as Lyman $\alpha$ forest studies.
The review of these issues was presented in
Ref.~\cite{BilGiuGriMas03}.

In this paper we study neutrino spin dynamics in medium with
stochastic characteristics such as number density, velocities and
polarizations of background fermions. The analysis is based on the
generalized relativistic invariant Bargmann-Michel-Telegdi
equation which describes neutrino spin evolution in presence of
general interactions. That is why the type of the neutrino
interactions with medium remains arbitrary. This equation enables
one to study the neutrino spin precession at arbitrary moving and
polarized matter even at the absence of the external
electromagnetic fields. This case can be of great importance if a
neutrino is treated, for example, within the context of minimally
extended standard model. The magnetic moment of such a neutrino
has quite small value (see, e.g.,
Refs.~\cite{FujShr80,DvoStu04PRD}),
\begin{equation}\label{NuMagnMom}
  \mu_\nu\approx 3.2\times 10^{-19}\mu_B
  \left(
    \frac{m_\nu}{\text{eV}}
  \right).
\end{equation}
The external electromagnetic field cannot effectively influence on
the spin dynamics of a neutrino with such electromagnetic
properties. We derive the most general equation for the
description of the averaged neutrino helicity evolution. It is
found that the neutrino helicity relaxation rate depends on the
correlation functions of the hydrodynamical currents of background
fermions. These correlators are calculated for the
ulrarelativistic electron-positron plasma. The results of the
paper are used for the evaluation of cosmological upper bound on
the $\tau$-neutrino mass.

We start with the generalized Bargmann-Michel-Telegdi equation for
the neutral $1/2$ spin fermion (e.g., neutrino) interacting with
arbitrary moving and polarized matter (see Ref.~\cite{LobStu01}),
\begin{align}
  \frac{d S^\mu}{d\tau} & = 2\mu
  (G^{\mu\nu}S_\nu-u^\mu G_{\alpha\beta}u_\alpha S_\beta)
  \notag
  \\
  \label{BMTgen}
  & +
  2\varepsilon
  (\widetilde{G}^{\mu\nu}S_\nu-
  u^\mu\widetilde{G}_{\alpha\beta}u_\alpha S_\beta),
\end{align}
where $G_{\mu\nu}=(-\mathbf{P},\mathbf{M})$ is the antisymmetric
tensor which accounts for the matter effects,
$\widetilde{G}_{\mu\nu}=
(1/2)\varepsilon_{\mu\nu\alpha\beta}G^{\alpha\beta}=
(\mathbf{M},\mathbf{P})$ is the tensor dual to $G_{\mu\nu}$, $\mu$
and $\varepsilon$ are the magnetic and electric dipole moments of
a particle, $u^\mu$ is the four velocity and $S^\mu$ is the four
dimensional spin vector. The derivative is taken with respect to
the particle's proper time $\tau$.

According to the results of Ref.~\cite{LobStu01} the tensor
$G_{\mu\nu}$ can be represented in the following way
\begin{equation*}
  G_{\mu\nu}=\varepsilon_{\mu\nu\alpha\beta}g^{\alpha}u^{\beta}-
  (f_{\mu}u_{\nu}-f_{\nu}u_{\mu}),
\end{equation*}
where the four vectors $g^{\mu}=(g^0,\mathbf{g})$ and
$f^{\mu}=(f^0,\mathbf{f})$ depend on the four vectors of
hydrodynamical currents and polarizations of medium fermions,
\begin{align*}
  g^{\mu} & =\sum_{f}(\rho^{(1)}_{f}j^{\mu}_f+
  \xi^{(1)}_{f}\lambda^{\mu}_f),
  \\
  f^{\mu} & =\sum_{f}(\rho^{(2)}j^{\mu}_f+
  \xi^{(2)}\lambda^{\mu}_f).
\end{align*}
The constants, $\rho^{(1,2)}_{f}$ and $\xi^{(1,2)}_{f}$, are fixed
by the model of the neutrino interaction with background fermions,
the sum is taken over all fermionic species. It is also possible
to represent the dependence of the vectors $\mathbf{M}$ and
$\mathbf{P}$ on the components of the vectors $g^{\mu}$ and
$f^{\mu}$ (see Ref.~\cite{LobStu01}),
\begin{align*}
  \mathbf{M} & =\gamma
  \left\{
    (g^0\bm{\beta}-\mathbf{g})-
    [\bm{\beta}\times\mathbf{f}]
  \right\},
  \\
  \mathbf{P} & =-\gamma
  \left\{
    (f^0\bm{\beta}-\mathbf{f})+
    [\bm{\beta}\times\mathbf{g}]
  \right\},
\end{align*}
where $\bm{\beta}$ is the neutrino velocity. The dependence of
these vectors on the velocities and polarizations of the
background fermions is also presented in Ref.~\cite{LobStu01} in
the explicit form.

It is convenient to rescale the vectors $\mathbf{M}$ and
$\mathbf{P}$ to exclude magnetic and electric dipole moments from
Eq.~\eqref{BMTgen},
\begin{equation*}
  \mathbf{M}\to\frac{\mathbf{M}}{\mu},
  \quad
  \mathbf{P}\to\frac{\mathbf{P}}{\varepsilon}.
\end{equation*}
Then we should rewrite Eq.~\eqref{BMTgen} using three dimensional
neutrino spin vector, $\bm{\zeta}$, which is related to the four
dimensional neutrino spin vector by the following formula
\begin{equation}\label{Szeta}
  S^\mu=
  \left(
    \frac{(\bm{\zeta}\mathbf{p})}{m_\nu},
    \bm{\zeta}+
    \frac{\mathbf{p}(\bm{\zeta}\mathbf{p})}{m_\nu(m_\nu+E_\nu)}
  \right),
\end{equation}
where $\mathbf{p}$ is the neutrino momentum. With help of
Eqs.~\eqref{BMTgen} and \eqref{Szeta} we obtain the equation for
the vector $\bm{\zeta}$,
\begin{equation}\label{zetaMP}
  \frac{d\bm{\zeta}}{d t}=
  \frac{2}{\gamma}[\bm{\zeta}\times\mathbf{M}_0]-
  \frac{2}{\gamma}[\bm{\zeta}\times\mathbf{P}_0].
\end{equation}
Note that the dynamics of the neutrino spin is determined by the
vectors $\mathbf{M}$ and $\mathbf{P}$ in the neutrino rest frame,
\begin{align*}
  \mathbf{M}_0 & =\gamma\bm{\beta}
  \left(
    g^0-\frac{1}{1+\gamma^{-1}}(\mathbf{g}\bm{\beta})
  \right)-
  \mathbf{g},
  \\
  \mathbf{P}_0 & =-\gamma\bm{\beta}
  \left(
    f^0-\frac{1}{1+\gamma^{-1}}(\mathbf{f}\bm{\beta})
  \right)+
  \mathbf{f}.
\end{align*}
It is convenient to introduce the new vector
$\bm{\Omega}=\mathbf{M}_0-\mathbf{P}_0$ and rewrite
Eq.~\eqref{zetaMP} in the form,
\begin{equation}\label{zetaOmega}
  \frac{d\bm{\zeta}}{d t}=
  \frac{2}{\gamma}[\bm{\zeta}\times\bm{\Omega}].
\end{equation}
The neutrino spin precession is now determined by the vector
$\bm{\Omega}$ which can be expressed in terms of the components of
the new auxiliary four vector $A^\mu=(A^0,\mathbf{A})$,
\begin{equation}\label{Omega}
  \bm{\Omega}=\gamma\bm{\beta}
  \left(
    A^0-\frac{1}{1+\gamma^{-1}}(\mathbf{A}\bm{\beta})
  \right)-
  \mathbf{A}.
\end{equation}
The four vector $A^\mu$ also depends on the four vectors of
hydrodynamical currents and polarizations of medium fermions,
\begin{equation*}
  A^\mu=g^\mu+f^\mu=\sum_{f}(\rho_{f}j^{\mu}_f+
  \xi_{f}\lambda^{\mu}_f),
\end{equation*}
where the coefficients $\rho_{f}$ and $\xi_{f}$ are related to the
constants $\rho^{(1,2)}_{f}$ and $\xi^{(1,2)}_{f}$,
\begin{equation*}
  \rho_f=\rho^{(1)}_{f}+\rho^{(2)}_{f},
  \quad
  \xi_f=\xi^{(1)}_{f}+\xi^{(2)}_{f}.
\end{equation*}

Now let us turn to the description of the neutrino helicity
evolution. Basing upon Eqs.~\eqref{zetaOmega} and \eqref{Omega}
one obtains the equation for particle's helicity,
$h=(\bm{\beta}\bm{\zeta})/\beta$, which has the form
\begin{equation}\label{helraw}
  \frac{d h}{d t}=
  \frac{2}{\gamma}(\mathbf{n}[\bm{\zeta}\times\bm{\Omega}])=
  -\frac{2}{\gamma}(\mathbf{A}[\mathbf{n}\times\bm{\zeta}]),
\end{equation}
where $\mathbf{n}=\bm{\beta}/\beta$ is the unit vector along the
neutrino velocity.

We can solve Eq.~\eqref{helraw} for the description of the
neutrino helicity evolution by means of iterations. The similar
method for the description of the neutrino helicity evolution in
stochastic electromagnetic fields was proposed in
Ref.~\cite{LoeSto89}. One substitutes the formal integral solution
of Eq.~\eqref{zetaOmega},
\begin{equation*}
  \bm{\zeta}(t)=\frac{2}{\gamma}
  \int_0^t\mathrm{d}t'[\bm{\zeta}(t')\times\bm{\Omega}(t')],
\end{equation*}
in Eq.~\eqref{helraw}. After some uncomplicated but rather
cumbersome calculations we obtain the equation for the neutrino
helicity,
\begin{align}\label{helA}
  \frac{d h}{d t} & =-
  \left(
    \frac{2}{\gamma}
  \right)^2
  \int_0^t\mathrm{d}t'
  \big\{
    (\mathbf{A}_\perp(\mathbf{r},t)
    \mathbf{A}_\perp(\mathbf{r}',t'))h(t')
    \notag
    \\
    &
    +
    \gamma(\bm{\zeta}_\perp(t')\mathbf{A}_\perp(\mathbf{r},t))
    \left[
      \beta A^0(\mathbf{r}',t')-A_\parallel(\mathbf{r}',t')
    \right]
  \big\}.
\end{align}
The longitudinal, $A_\parallel=(\mathbf{n}\mathbf{A})$, and
transversal,
$\mathbf{A}_\perp=\mathbf{A}-\mathbf{n}(\mathbf{n}\mathbf{A})$,
components of the vector $\mathbf{A}$ are taken with respect to
the neutrino velocity. The vectors $\mathbf{r}$ and $\mathbf{r}'$
are the neutrino positions at the times $t$ and $t'$. If we
suppose that a neutrino propagates along a line, then
$\mathbf{r}'-\mathbf{r}=\bm{\beta}(t'-t)$.

Now let us discuss the average quantities in Eq.~\eqref{helA} that
will be denoted as $\langle\dots\rangle$. We have already
mentioned that the neutrino helicity evolution is suppressed by
the factor $1/\gamma$ which is small for relativistic neutrinos.
This feature can be also derived from Eq.~\eqref{helA}. Thus,
if we consider Eq.~\eqref{helA} to the lowest order in $1/\gamma$,
the variation of the averaged neutrino helicity turns out to be
negligibly small during the correlation time of the vector $A^\mu$
components, because the characteristic time scale of the
correlators $\langle
A^\mu(\mathbf{r},t)A^\nu(\mathbf{r}',t')\rangle$ for the high
temperature plasma is very short. Let us evaluate this time scale
of the correlators when a $\tau$-neutrino interacts with high
temperature electron-positron plasma (this problem will be
discussed in details below). The time scale is approximately equal
to the mean free path of plasma particles (see, e.g.,
Ref.~\cite{Akh74p37}), i.e. electrons and positrons,
\begin{equation*}
  \mathfrak{t}\approx \frac{T^2}{\alpha^2_\mathrm{em}n\Lambda},
\end{equation*}
where $T$ is the plasma temperature, $n$ in the number density in
plasma, $\alpha_\mathrm{em}=e^2$ is the fine structure constant
and
\begin{equation*}
  \Lambda=\ln
  \left(
    \frac{T}{\alpha_\mathrm{em}}
    \sqrt{\frac{T}{4\pi\alpha_\mathrm{em}n}}
  \right),
\end{equation*}
is the Coulomb logarithm.

Supposing that for the relativistic plasma one has $n\approx
0.36T^3$ (here we take into account that both electrons and
positrons conribute to the number density) and setting $T\approx
100\thinspace\text{MeV}$, we get for the mean free path the
following estimate, $\mathfrak{t}\approx 5.2\times
10^{-20}\thinspace\text{s}$. On the other hand one obtains for the
typical time of the averaged neutrino helicity relaxation (see
Eq.~\eqref{varrhoE} below), $1/\Gamma\approx 8.6\times
10^{-4}\thinspace\text{s}$. One can see from this estimates that
$\mathfrak{t}\ll 1/\Gamma$. That is why we can consider the
integrand in Eq.~\eqref{helA} as the product of two functions: the
correlators of the vector $A^\mu$ components and the averaged
neutrino helicity.

The vector $A^\mu$ is the linear combination of the
hydrodynamical currents and polarizations vectors. Different
components of these vectors are sure to be uncorrelated. Thus we
get that the correlators of different components of the vector
$A^\mu$ vanish. Finally one obtains more simple equation for the
averaged neutrino helicity
\begin{equation}\label{hfint}
  \frac{d \langle h \rangle}{d t}=-
  \left(
    \frac{2}{\gamma}
  \right)^2
  \int_0^t\mathrm{d}t'
  \langle
    \mathbf{A}_\perp(\mathbf{r},t)\mathbf{A}_\perp(\mathbf{r}',t')
  \rangle
  \langle
    h(t')
  \rangle.
\end{equation}
Eq.~\eqref{hfint} is the most general one which describes the
averaged neutrino helicity evolution in presence of matter with
random velocities and polarizations. Note that the type of the
neutrino interaction with background fermions is not fixed yet.

Eq.~\eqref{hfint} can be transformed to the more convenient form
which will be used in the further analysis. As we have already
noticed the correlations of the fields $\mathbf{A}$ are not equal
to zero for a rather short period of time. Thus supposing again
that the neutrino helicity is practically constant during the
matter fields correlation period as well as taking into account
time-translation invariance of the matter fields correlators we
receive more simple differential equation for the mean neutrino
helicity
\begin{equation*}
  \frac{d \langle h \rangle}{d t}=
  -\Gamma \langle h \rangle,
\end{equation*}
where neutrino helicity relaxation rate has the form
\begin{equation}\label{GammaA}
  \Gamma=
  \left(
    \frac{2}{\gamma}
  \right)^2
  \int_0^\infty\mathrm{d}t
  \left.
    \langle
      \mathbf{A}_\perp(\mathbf{r},t)\mathbf{A}_\perp(0,0)
    \rangle
  \right|_{\mathbf{r}=\bm{\beta}t}.
\end{equation}
We can take into account only
$\langle\mathbf{j}_{f}(\mathbf{r},t)\mathbf{j}_{f}(0,0)\rangle$
correlation functions in Eq.~\eqref{GammaA} if we consider
neutrino interacting with the relativistic plasma. Indeed for
relativistic fermions the polarization vector is proportional to
the current, $\bm{\lambda}_f\approx h_f\mathbf{j}_f$, where
$h_f=(\bm{\beta}_f\bm{\zeta}_f)/\beta_f$ is the fermion helicity,
$\bm{\beta}_f$ is the velocity of the reference frame in which the
mean momentum of $f$-type background fermions is zero and
$\bm{\zeta}_f$ is the mean value of the polarization vector of the
$f$-type background fermions in the above mentioned reference
frame. The fermion helicities and currents in a stochastic plasma
take the form
\begin{equation*}
  h_f=\langle h_f \rangle+\delta h_f,
  \quad
  \mathbf{j}_f=\langle \mathbf{j}_f \rangle+\delta\mathbf{j}_f,
\end{equation*}
where $\langle h_f \rangle=0$ and $\langle \mathbf{j}_f \rangle=0$
are the mean values of the corresponding quantities. The
fluctuations of fermion helicities, $\delta h_f$, and currents,
$\delta\mathbf{j}_f$, are taken to be small. Therefore the
correlation functions like
$\langle\bm{\lambda}_{f}(\mathbf{r},t)\mathbf{j}_{f}(0,0)\rangle$
and
$\langle\bm{\lambda}_{f}(\mathbf{r},t)\bm{\lambda}_{f}(0,0)\rangle$
are next to the leading order.

The hydrodynamical currents correlation functions can be expressed
in terms of the electromagnetic currents correlators for plasma
consisted of charged particles. If the charges of particles in
plasma are equal to $\pm e$, then
\begin{equation*}
  \langle
    \mathbf{j}_{f}(\mathbf{r},t)\mathbf{j}_{f}(0,0)
  \rangle=
  \frac{1}{e^2}
  \langle
    \mathbf{j}^\mathrm{em}_{f}(\mathbf{r},t)
    \mathbf{j}^\mathrm{em}_{f}(0,0)
  \rangle,
\end{equation*}
where $\mathbf{j}^\mathrm{em}_{f}(\mathbf{r},t)$ is the
electromagnetic current. Thus the neutrino helicity relaxation
rate is expressed in the following way
\begin{align}
  \Gamma & =
  \left(
    \frac{2}{\gamma}
  \right)^2
  \frac{1}{e^2}
  \sum_f\rho^2_f
  \notag
  \\
  \label{Gammajem}
  & \times
  \int_0^\infty\mathrm{d}t
  \left.
    \langle
      \mathbf{j}^\mathrm{em}_{f\perp}(\mathbf{r},t)
      \mathbf{j}^\mathrm{em}_{f\perp}(0,0)
    \rangle
  \right|_{\mathbf{r}=\bm{\beta}t}.
\end{align}

The correlation functions in Eq.~\eqref{Gammajem} are usually
expressed in terms of the Fourier transforms,
\begin{multline}\label{jjFour}
  \langle
    j^\mathrm{em}_{i}(\mathbf{r},t)
    j^\mathrm{em}_{j}(0,0)
  \rangle
  \\
  =
  \frac{1}{(2\pi)^4}
  \int\mathrm{d}\omega\mathrm{d}^3\mathbf{k}
  (j^\mathrm{em}_{i}j^\mathrm{em}_{j})_{\omega\mathbf{k}}
  e^{-i\omega t+i\mathbf{k}\mathbf{r}}.
\end{multline}
The electromagnetic currents correlators can be obtained with help
of the fluctuation-dissipation theorem (see Ref.~\cite{CalWel51}).
The explicit form of the correlation functions is presented in
Ref.~\cite{Akh74p518},
\begin{widetext}
\begin{equation}
  (j^\mathrm{em}_i j^\mathrm{em}_j)_{\omega\mathbf{k}} 
  =
  \frac{\omega^2}{2\pi}
  \frac{1}{\exp(\omega/T)-1}
  \label{jjexpl}
  \times
  \left\{
    \frac{k_i k_j}{k^2}
    \frac{\Im\mathfrak{m}\epsilon_l}{|\epsilon_l|^2}+
    \left(
      \delta_{ij}-
      \frac{k_i k_j}{k^2}
    \right)
    \left(
      1-(k/\omega)^2
    \right)^2
    \frac{\Im\mathfrak{m}\epsilon_t}{|\epsilon_t-(k/\omega)^2|^2}
  \right\}.
\end{equation}
The correlators are expressed in terms of the permittivity tensor
components, $\epsilon_l$ and $\epsilon_t$. For the case of the
ultrarelativistic plasma, $T\gg m_f$, where $m_f$ is the
background fermion mass, the longitudinal and transversal
components of the permittivity tensor take the form (see, e.g.,
Ref.~\cite{LifPit79p165})
\begin{align}
  \label{epsl}
  \epsilon_l(\mathbf{k},\omega) & =
  1+\frac{1}{a^2 k^2}
  \left\{
    1+\frac{\omega}{2k}
    \ln
    \left[
      \frac{|k-\omega|}{k+\omega}
    \right]+
    i\frac{\pi\omega}{4k}
    [1+\mathrm{sign}(k-\omega)]
  \right\},
  \\
  \epsilon_t(\mathbf{k},\omega) & =
  1-\frac{1}{2a^2 k^2}
  \label{epst}
  \times
  \left\{
    1-\frac{k}{2\omega}
    \left(
      1-(\omega/k)^2
    \right)
    \ln
    \left[
      \frac{|k-\omega|}{k+\omega}
    \right]-
    i\frac{\pi k}{4\omega}
    \left(
      1-(\omega/k)^2
    \right)
    [1+\mathrm{sign}(k-\omega)]
  \right\},
\end{align}
\end{widetext}
where $a$ is the Debye length which is related to the particles
densities in plasma by the following formula
\begin{equation*}
  a^{-2}=
  \frac{4\pi e^2}{T}\sum_f n_f,
\end{equation*}
and
\begin{equation*}
  \mathrm{sign}(x)=
  \begin{cases}
    1, & x>0, \\
    -1, & x<0.
  \end{cases}
\end{equation*}
is the step function.

As it was pointed out in Ref.~\cite{LoeSto89}, we can take into
account only longitudinal component of the electromagnetic
currents correlators since the imaginary part of the transversal
component of the permittivity tensor [which contributes to the
transversal component of the electromagnetic currents correlators,
see Eq.~\eqref{jjexpl}] is suppressed by the additional small
factor $1/\gamma^2$. Using Eqs.~\eqref{Gammajem}-\eqref{epst} one
obtains for the neutrino helicity relaxation rate,
\begin{widetext}
\begin{align}
  \Gamma & \approx
  \left(
    \frac{2}{\gamma}
  \right)^2
  \frac{1}{e^2}
  \sum_f\rho^2_f
  \frac{\beta}{16\pi^2a^2}
  \int_{-1}^{+1}
  \mathrm{d}(\cos\vartheta)\sin^2\vartheta\cos\vartheta
  \int_0^\infty\mathrm{d}k
  \left.
    \frac{\omega^2}{\exp(\omega/T)-1}
    \frac{1}{|\epsilon_l|^2}
  \right|_{\omega=k\beta\cos\vartheta}
  \notag
  \\
  \label{GammaN}
  & \approx
  \left(
    \frac{2}{\gamma}
  \right)^2
  \frac{1}{e^2}
  \sum_f\rho^2_f
  \frac{T^5}{12\pi^2(aT)^2}
  \int_0^\infty\mathrm{d}z
  \frac{z^2}{e^z-1}\approx
  0.19 N T^5
  \frac{m_\nu^2}{E_\nu^2}
  \sum_f\rho^2_f,
\end{align}
\end{widetext}
where $N$ is the number of particles species in plasma and
$\vartheta$ is the angle between vectors $\mathbf{k}$ and
$\bm{\beta}$. In Eq.~\eqref{GammaN} we drop small terms like
$1/\gamma^4$ as well as we keep the leading terms in the fine
structure constant. We also suppose the frequency $\omega$ and the
wave vector $\mathbf{k}$ are connected by the relation,
$\omega=(\mathbf{k}\bm{\beta})=k\beta\cos\vartheta$, that
immediately follows from Eq.~\eqref{Gammajem}.

The neutrino helicity relaxation due to the neutrino magnetic
moment interaction with the stochastic electromagnetic field was
studied in Ref.~\cite{LoeSto89}. The helicity relaxation rate
obtained in that paper is proportional to $\mu^2$, where $\mu$ is
the neutrino magnetic moment. One can also consider the
correlations between stochastic electromagnetic fields and
polarizations of the background fermions. Indeed the background
fermions are polarized under the influence of the external
electromagnetic field. In this case the induced polarization of
the $f$-type fermions has the form,
\begin{equation*}
  \bm{\zeta}^{(\mathrm{ind})}_f=\frac{1}{n_f}
  \left(
    \frac{1}{g^{(e)}_f}\kappa_f\mathbf{E}+
    \frac{1}{g^{(m)}_f}\frac{\chi_f}{1+\chi_f}\mathbf{B}
  \right),
\end{equation*}
where $\kappa_f$ and $\chi_f$ are dielectric and magnetic
susceptibilities, $g^{(e)}_f$ and $g^{(m)}_f$ are the gyroelectric
and gyromagnetic ratios, $n_f$ is the number density of the
$f$-type background fermions. In the latter case the corresponding
contribution to the neutrino helicity relaxation rate will be
proportional to the factor $\mu\times\gamma^{-1}$.

Supposing that neutrino magnetic moment is described within a
realistic model, for instance the minimally extended standard
model (see Refs.~\cite{FujShr80,DvoStu04PRD}), we obtain, e.g.,
for a $\tau$-neutrino (this neutrino type will be considered
below) magnetic moment $\mu_{\nu_\tau}\approx 5.8\times
10^{-12}\mu_B$ [see also Eq.~\eqref{NuMagnMom}]. The
$\tau$-neutrino mass is taken to be $18.2\thinspace\text{MeV}$
\cite{Eid04}. If a particle has such a small magnetic moment, the
contribution of the corresponding terms to the neutrino helicity
relaxation rate will be negligibly small compared to the
contribution examined in our paper.

To evaluate the mean neutrino energy in Eq.~\eqref{GammaN} we note
that it can be expressed with help of the neutrino energy and
number densities $\langle E_\nu \rangle=\varrho_E/n_\nu$. For the
relativistic neutrino gas which is in equilibrium at the
temperature $T$ we have
\begin{equation}\label{varrhoE}
  \varrho_E=2\int
  \frac{\mathrm{d}^3\mathbf{p}}{(2\pi)^3}
  \frac{E_\nu}{\exp(E_\nu/T)+1}\approx 0.58 T^4,
\end{equation}
and
\begin{equation}\label{nnu}
  n_\nu=2\int
  \frac{\mathrm{d}^3\mathbf{p}}{(2\pi)^3}
  \frac{1}{\exp(E_\nu/T)+1}\approx 0.18 T^3.
\end{equation}
Finally we get for the mean neutrino energy $\langle E_\nu
\rangle\approx 3.22 T$.

Let us discuss a $\tau$-neutrino propagating in ultrarelativistic
plasma composed of electrons and positrons. This neutrino flavor
eigenstate interacts with electron-positron plasma by means of the
weak neutral currents. In this case the coefficients $\rho_f$ are
expressed in the following way
\begin{equation}\label{coeff}
  \rho_{e^-}=-\rho_{e^+}=
  -\frac{G_F}{2\sqrt{2}}(1-4\sin^2\theta_W),
\end{equation}
where $G_F$ is the Fermi constant and $\theta_W$ is the Weinberg
angle. Using Eqs.~\eqref{GammaN}-\eqref{coeff}
we obtain for the neutrino helicity relaxation rate,
\begin{equation}\label{GammaT}
  \Gamma\approx 1.16\times 10^{3}
  \left(
    \frac{m_{\nu_\tau}}{10\thinspace\text{MeV}}
  \right)^2
  \left(
    \frac{T}{100\thinspace\text{MeV}}
  \right)^3\thinspace\text{s}^{-1}.
\end{equation}

The main results of the paper can be applied, e.g., for the
description of the neutrino spin relaxation in the primordial
plasma. The number of right-handed neutrinos is very small at
present. Indeed if one observed a right-handed neutrino, it would
be a direct indication on the neutrino having a non-zero mass
since the wave function of a massive particle can correspond to
both left- and right-handed helicity states contrary to the case
of a massless particle. However, up to now only the upper bound on
the neutrino mass is experimentally established. Thus we should
assert that if there were right-handed neutrinos, their number
would be not very great. The neutrinos with this polarization
cannot be produced in great amount at the early stages of the
Universe evolution. Thus we should suppose that the process of
neutrino helicity relaxation (i.e. the production of right-handed
neutrinos since all neutrinos are taken to be left-handed
initially) is out of the thermodynamical equilibrium, i.e.
$\Gamma<H$, where $H$ is the Hubble constant. At radiation
dominated era of the early Universe the Hubble constant can be
approximately evaluated (see, for instance,
Ref.~\cite{Wei72p534}),
\begin{equation}\label{Hubble}
  H\approx(G_N)^{1/2}T^2\approx 1.24\times 10^{3}
  \left(
    \frac{T}{100\thinspace\text{MeV}}
  \right)^2\thinspace\text{s}^{-1},
\end{equation}
where $G_N$ is the Newton's constant.

When temperature of the early Universe is equal or less than
$100\thinspace\text{MeV}$, the muons and antimuons densities are
negligible to effectively contribute to the neutrino helicity
relaxation. The densities of neutrons and protons are also small.
Thus a $\tau$-neutrino can interact only with electrons and
positrons at this temperature. From Eqs.~\eqref{GammaT} and
\eqref{Hubble} for $T=100\thinspace\text{MeV}$ we get
$m_{\nu_\tau}<10.2\thinspace\text{MeV}$. The obtained upper limit
on the $\tau$-neutrino mass is compatible with the up to date
neutrino mass bounds derived in various techniques (see
Ref.~\cite{Eid04}). It should be noted that the calculations made
in this paper are also consistent with the obtained upper limit.
Indeed in deriving of Eq.~\eqref{GammaN} we supposed that
$1/\gamma^4=(m_\nu/E_\nu)^4\ll 1$. Eqs.~\eqref{varrhoE} and
\eqref{nnu} are valid while $m_\nu<T$. All these assumptions are
in agreement with the temperature of the primordial plasma
($T=100\thinspace\text{MeV}$) and bound on the neutrino mass
($m_{\nu_\tau}\lesssim 10\thinspace\text{MeV}$). It is worth
mentioning that the decoupling temperature of $\tau$-neutrinos
found in Ref.~\cite{Han04} is $\approx 3.7\thinspace\text{MeV}$.
Thus down to this temperature $\tau$-neutrinos are in thermal
equilibrium with primeval plasma.

Although we studied the $\tau$-neutrino helicity relaxation in the
early Universe, the interaction of the neutrino spin with the
gravitational field (space-time curvature) was not taken into
account in our work, i.e. all the calculations were performed in
the minkowskian space-time. However the neutrino propagation in
strong gravitational fields of the specific configurations (e.g.,
Kerr metric) can cause the asymmetry between neutrinos and
antineutrinos. The most recent analysis of this issue is presented
in Ref.~\cite{Muk05}. In spite of the fact that the Universe is
assumed to expand isotropically this effect is of great importance
on the early stages of the Universe evolution when relic neutrinos
interact with primordial black holes before neutrinos decoupled
(see Ref.~\cite{SinMuk03}). This process could influence the
neutrino helicity relaxation.

It should be noted that for the first time the cosmological upper
bound on the sum of all neutrinos masses was derived in
Ref.~\cite{GerZel66}. One can obtain the contribution of stable
light neutrinos to the matter density in the Universe and then
compare the calculated value of the neutrino contribution with the
observational value of total matter density. The "classical"
cosmological bound (Gerstein-Zeldovich limit) is very small
\begin{equation*}
  \sum_i m_{\nu_i}<14\thinspace\text{eV}.
\end{equation*}
In our paper the production rate of "wrong-helicity" (i.e.
right-handed) Dirac neutrinos was constrained by the Hubble
parameter. For the first time the similar idea was used in
Ref.~\cite{ShaTeuWas80}. It was assumed that if the population of
the right-handed neutrino states in the primordial plasma is
great, then they could influence the primordial nucleosynthesis by
enlarging the effective number of the neutrino species.

In conclusion we note that the helicity evolution of a neutrino
interacting with matter having random distributions of number
density, velocities and polarizations has been studied in this
paper. The basic equation was the generalized relativistic
invariant Bargmann-Michel-Telegdi equation. We have derived the
expression for the neutrino helicity relaxation rate which depends
on the correlation functions of the hydrodynamical currents of
background fermions. These correlators have been evaluated for the
case of the ultrarelativistic electron-positron plasma. Then we
have studied $\tau$-neutrino helicity evolution in plasma of the
early Universe at $T=100\thinspace\text{MeV}$. Comparing the
helicity relaxation rate with the Hubble constant we have obtained
the upper bound on the $\tau$-neutrino mass which turned out to be
$\approx 10\thinspace\text{MeV}$. This $\tau$-neutrino mass limit
is consistent with modern laboratory constraints.

\begin{acknowledgments}
This research was supported by grants of Deutscher Akademischer
Austausch Dienst and Russian Science Support Foundation.
\end{acknowledgments}

\bibliography{generaleng}

\end{document}